\documentclass[a4paper,11pt]{article}
\usepackage{jheppub} 
\usepackage{lineno}

\usepackage{amsmath,amssymb,mathtools} 
\usepackage{color}
\usepackage{graphicx}
\usepackage{subfigure}
\usepackage{caption}        
\usepackage{hyperref}            
\usepackage{multirow,makecell}   
\usepackage{textcomp}
\usepackage{wasysym}
\usepackage{setspace}
\usepackage{verbatim}
\usepackage{float}
\usepackage{ulem}
\usepackage{mhchem}
\usepackage[utf8]{inputenc}
\usepackage{bm}

\title{Splitting dynamics of quantized composite vortices in holographic miscible binary superfluids}

\author[a,b]{Yu-Ping An}%
\author[a,b,c]{Li Li}
 \affiliation[a]{CAS Key Laboratory of Theoretical Physics, Institute of Theoretical Physics,
Chinese Academy of Sciences, Beijing 100190, China}
\affiliation[b]{School of Physical Sciences, University of Chinese Academy of Sciences, Beijing 100049, China}
\affiliation[c]{School of Fundamental Physics and Mathematical Sciences, Hangzhou Institute for Advanced Study, University of Chinese Academy of Sciences, Hangzhou 310024, China}
\emailAdd{anyuping@itp.ac.cn, liliphy@itp.ac.cn}

\abstract{
The stability properties and splitting dynamics of multiply quantized vortices are the subject of interest in both theoretical and experimental investigations. Going beyond the regime of validity of Gross-Pitaevskii equation (GPE), we study the composite vortices in miscible strongly interacting binary superfluids by employing a holographic model that naturally incorporate finite temperature and dissipation. The composite vortices is classified in terms of an integer pair $(S_1, S_2)$ of phase winding numbers and can share the same vortex core, while either co-rotating or counter-rotating, leading to very diverse vortex structures. We uncover different dynamical behaviors compared to results from GPE that is valid in weak coupling limit and zero temperature. In particular, we show that the occurrence of dynamic instabilities and the instability strength are sensitive to the temperature. We identify several temperature dependent dynamical transitions in $(1,1)$, $(2,\pm 1)$ and $(2,2)$ vortices. The splitting behaviors associated with different multipolarities are demonstrated by solving the full-time evolution for slightly perturbed composite vortices. We find that the final states of all composite vortices are generally singly quantized vortices, and no additional long living vortex is formed due to strong dissipation. Our results highlight the important role of temperature and the distinction between dynamics of composite vortices in weakly interacting superfluids without dissipation and strongly interacting case with dissipation, shedding a new light on the understanding of quantum vortex and dynamical instabilities in multicomponent superfluids.
}

\begin{document}
\maketitle
\flushbottom

\section{Introduction}

As a macroscopic quantum phenomenon, superfluidity is characterized by the flow of matter without resistance and has been observed in various systems. Except at absolute zero temperature, a superfluid consists of two inseparable fluids-normal fluid and superfluid component. The latter has no friction, dissipation, or viscosity. A major part of understanding the behavior of quantum fluids is understanding the interaction between the superfluid component and the normal component, such as effects of dissipation and defects. The quantum nature of superfluidity leads to quantum vortices that are topological defects and manifest as quantized whirlpools of circulation. Each vortex is characterized by a singular core where the phase of the superfluid order parameter winds by $2\pi S$, with the winding number $S$ an integer. The topological nature of these vortices significantly affects the possible flow patterns of quantum fluids. Theoretical descriptions of vortex dynamics in superfluid systems have progressed significantly through a blend of analytical insights and numerical simulations. The mean-field Gross-Pitaevskii equation (GPE) provides a theoretic framework for describing the dynamics of vortices in the weakly coupled case in the absence of any normal fluid component. Meanwhile, recent advances in holographic superfluids offer a powerful way to explore the strong coupling regime, incorporating naturally finite temperature and dissipation effects~\cite{PhysRevLett.101.031601,Herzog:2008he}.

A quantum vortex is a localized excitation that retains its shape over time. Nevertheless, if the vortex is not stable under small perturbations, then in any real-life situation, the vortex will not retain its configuration. Multiply quantized vortices with the winding number $|S|>1$ generally present splitting instability, both in single-component and binary superfluids. Dynamical splitting instability of vortices in single component case has been widely studied, both in weakly coupled Bose-Einstein condensates (BECs)~\cite{RN400,RN399,RN352,RN401} and strongly coupled holographic superfluids~\cite{RN359,RN357,RN356,RN393,RN391,RN358}. The exploration of vortices in two-component superfluids has been growing, as it offers a rich opportunity to discover new physical phenomena and understand the interaction between different quantum phases~\cite{RN349,RN395,RN361,RN397,RN350,RN351,RN365,RN394,RN348,RN412,RN411,RN398}. In particular, exotic vortex configurations in binary BECs are within reach of state-of-the-art experiments.

The addition of a second superfluid component complexes the splitting instability of multiply quantized vortices. 
The dynamical instabilities and splitting of quantized composite vortices in two-component BECs have been studied by GPE~\cite{RN398,RN397}, which applies to weakly coupling regime and takes no account of normal fluid component. In this limit, composite vortices were found to present additional unstable modes. Nevertheless, it is important to consider the more realistic situation that incorporates finite temperature and dissipation. The holographic duality provides a useful framework by mapping the system into gravitational dynamics in AdS spacetime with one higher dimension, where finite temperature, finite density states are modeled by a black hole in the bulk. In our recent work~\cite{vortex-soliton}, we have studied vortex-birght soliton structure in strongly coupling immiscible binary superfluids, where the second component can exist in the core of the vortex in the first component and could act as a stabilizer. Such stabilization mechanism opens the possibility for vortices with smaller winding number to merge into vortices with larger winding number, which was confirmed for the first time in our simulation. In the present work, we shall consider the composite vortices in miscible strongly interacting binary superfluids, going beyond the regime of validity of GPEs by employing a holographic binary superfluid model. In this case, quantum vortices in two components can share the same vortex core while having different winding numbers. We will show numerically the dynamical instabilities and splitting of singly and doubly quantized composite
vortices at finite temperature and dissipation. Some novel features will be uncovered compared with the zero-temperature GPEs. In particular,  our results demonstrate that the temperature effect as well as dissipation plays an important role in the the stability properties and splitting dynamics of multicomponent superfluids, which should therefore be amenable to experimental verification.

This paper is organized as follows. We describe our holographic model and setup in Section~\ref{sec2}. In Section~\ref{sec3}, we construct the stationary configurations of composite vortices in miscible binary superfluids and describe some properties. In Section~\ref{sec4}, we analyze the dynamical instabilities of composite vortices at different temperatures and coupling strengths using linear response theory. Then we explore the nonlinear regime of their dynamical instabilities and spitting patterns by full nonlinear time evolution. We conclude in Section~\ref{sec5} with some discussions.

\section{Holographic model and setup}
\label{sec2}
We start by establishing the gravitational description of a two-component superfluid in two spatial dimensions. We work in the probe limit by neglecting the back-reaction of matter content to the geometry. The action for matter content reads
    \begin{equation}
        \begin{aligned}
             S_m=&\int dx^4\sqrt{-g}[-(\mathcal{D}_\mu\Psi_1)^*
            \mathcal{D}^\mu\Psi_1-m_1^2|\Psi_1|^2-(\mathcal{D}_\mu\Psi_2)^*
            \mathcal{D}^\mu\Psi_2-m_2^2|\Psi_2|^2\\&-\frac{\nu}{2}|\Psi_1|^2|\Psi_2|^2-\frac{1}{4}F^{\mu\nu}F_{\mu\nu}]\,,
          \end{aligned}  
    \end{equation}
where $\mathcal{D}_\mu\Psi_i=(\nabla_\mu-ie_iA_\mu)\Psi_i$, $A_\mu$ is the $U(1)$ gauge field with $F_{\mu\nu}$ its strength. The two bulk scalars $\Psi_1$ and $\Psi_2$ are charged under the same $U(1)$ gauge field and represent the two superfluid components of the dual system. Early studies on holographic binary orders can be found \emph{e.g.} in~\cite{Basu:2010fa,Cai:2013wma,Yang:2019ibe,Yao:2022fov}. The intercomponent coupling strength $\nu$ characterizes the interaction between the two superfluid components and determines their miscibility. For $\nu>0$ this model describes immiscible binary superfluids, where the interface forms between the two components. Interface dynamics~\cite{An:2024ebg,An:2024dkn} and vortex-soliton structure~\cite{vortex-soliton} have been studied recently, for which some novel features were identified. In this work, we focus on $\nu<0$, \emph{i.e.} miscible binary superfluids. In~\cite{An:2024ctq} we have explored hydrodynamics and dynamical instability in homogeneous miscible binary superfluids. We showed that both the counterflow and coflow instabilities in binary superfluids are all essentially thermodynamic. 
Here we will study the dynamics of composite vortices in miscible binary superfluids.

In the probe limit, the fluctuations of the temperature and the normal fluid velocity are frozen, and meanwhile, the dynamics of momentum and energy are decoupled from the charge sector. The background metric is given by the Schwarzschild AdS black brane:
    \begin{equation}\label{backg}
        ds^2=\frac{L^2}{z^2}(-f(z)dt^2-2dtdz+dr^2+r^2d\theta^2),\; f(z)=1-\frac{z^3}{z_h^3}\,,
    \end{equation}
where $z=0$ denotes the AdS boundary and $z=z_h$ is the location of the event horizon. It corresponds to a thermal bath at temperature $T=3/(4\pi z_h)$ on the boundary. Note that we have used polar coordinates $(r, \theta)$ in spatial directions since we consider quantum vortices. For simplicity, we set $L=z_h=1$ and adopt the radial gauge $A_z=0$. We consider two identical components, specified by $m_1^2=m_2^2=-2$, $e_1=e_2=1$, corresponding to dual scalar operators with the scaling dimension $\Delta=2$.

Upon solving the equations of motion (EoMs) in the bulk, all relevant observables can be identified using the standard holographic dictionary, see~\cite{An:2024ebg,An:2024dkn}. Particularly near the AdS boundary $z=0$, their asymptotic expansions can be expressed as
    \begin{equation}\label{UV}
        \begin{aligned}
        A_\mu&=a_\mu+b_\mu z+\mathcal{O}(z^2),\quad
        \Psi_i=\Psi_i^{(v)} z^2+\mathcal{O}(z^3),\quad i=1,2\,.
        \end{aligned}
    \end{equation}
To break the $U(1)$ symmetry spontaneously, we have already turned off the leading source term of each scalar field. The sun-leading term $\Psi_i^{(v)}$ corresponds to the superfluid condensate $\mathcal{O}_i $. Moreover, $a_t=\mu$ is the chemical potential and $-b_t=\rho$ is the charge density. And $\bm{a}=(a_r, a_\theta)$ are related to the superfluid velocity $\bm{v}_i^s=\nabla\theta_i-\bm{a}$ with $\theta_i$ the phase of the superfluid condensation $\mathcal{O}_i$. For simplicity, we denote vectors in the boundary spatial directions using bold-face letters. We set $\bm{a}=0$, such that the superfluid velocity is given by $\bm{v}_i=\nabla\theta_i$.

Due to the scaling symmetry~\footnote{Here $(x, y)$ are Cartesian coordinates with $x=r\cos(\theta)$ and $y=r\sin(\theta)$.}
\begin{equation}
(t,x,y,z)\rightarrow \lambda(t,x,y,z),\quad (T,\mu,\bm{v}_i^s)\rightarrow \frac{1}{\lambda}(T,\mu, \bm{v}_i^s),\quad (\rho,\mathcal{O}_i)\rightarrow \frac{1}{\lambda^2}(\rho,\mathcal{O}_i)\,,
\end{equation}
with $\lambda$ a constant, $T$ and $\mu$ are not independent quantities. Since we have fixed $z_h=1$ (\emph{i.e.} $T=3/4\pi$), we choose $\mu$ to be a free parameter. There is a second-order phase transition occurring when $\mu \ge \mu_c \simeq 4.064$, assuming $z_h=1$. This condition also determines the ratio $T/T_c=\mu_c/\mu$ where $T_c$ is the critical temperature below which the superfluid phase develops spontaneously.

\section{Stationary composite vortex solutions}
\label{sec3}
We focus on the instabilities and
splitting of axisymmetric vortex states. To this end, we seek stationary solutions with the following ansatz in the bulk:
\begin{equation}
      \Psi_i=z\psi_i(z,r)e^{iS_i\theta+i\Theta(z,r)},\quad  A_t=A_t(z,r), \quad A_\theta=A_\theta(z,r),  
\end{equation}
where $S_i$ is the winding number of quantized vortex in the $i$-th component. In this work, we consider $S_1=1$, 2 and $-S_1\le S_2\le S_1$, to compare with weakly coupling results~\cite{RN398} based on the zero-temperature GPE. The phase $\theta_i$ in component $i$ is obtained from $(S_i\theta+\Theta)|_{z=0}$. For later convenience, we use $(S_1,S_2)$ vortex to refer the composite vortex with winding number $S_1$ in the first component and $S_2$ in the second component. Following~\cite{RN398}, we define a ``S-quantum composite vortex" as a $(S_1,S_2)$ vortex for which $\max_i |S_i|=S$. A ``coreless vortex" in
turn means either a $(S_1, 0)$ or $(0, S_2)$ vortex for which the total condensation $|\mathcal{O}_1|+|\mathcal{O}_2|$ does not vanish at
the phase singularity. In contrast, a vortex for which $\mathcal{O}_1=\mathcal{O}_2=0$ at the singularity is classified as ``cored."

The resulted EoMs are given as
\begin{equation}
\label{eompsi}
\begin{aligned}
    \partial_z(f\partial_z\psi_i)+\partial_r^2\psi+\frac{1}{r}\partial_r\psi+(\frac{A_t^2}{f}-\frac{(A_\theta-S_i)^2}{r^2}-z-\frac{\nu}{2}\psi_j^2)\psi_i=0, \\(i,j=1,2,\quad i\ne j)\,,
\end{aligned}
\end{equation}
\begin{equation}
\label{eomat}
    f\partial_z^2A_t+\partial_r^2A_t+\frac{1}{r}\partial_rA_t-2A_t\sum_i\psi_i^2=0\,,
\end{equation}
\begin{equation}
\label{eomatheta}
    \partial_z(f\partial_zA_\theta)+\partial_rA_\theta-\frac{1}{r}\partial_rA_\theta-2\sum_i(A_\theta-S_i)\psi_i^2=0\,.
\end{equation}
Note that the EoMs for $\Theta$ and $A_r$ have be eliminated by setting $\partial_z\Theta=-A_t/f$ and $A_r=\partial_r\Theta$. The above coupled nonlinear ordinary differential equations are solved with the following boundary conditions. At the AdS boundary $z=0$, one has
\begin{equation}
\begin{aligned}
    \psi_i|_{z=0}=0, \quad A_t|_{z=0}=\mu, \quad A_\theta|_{z=0}=0.
\end{aligned}    
\end{equation}

We consider the regular boundary conditions at the event horizon, in particular $A_t|_{z=z_h}=0$. In the $r$ direction, the
system is cut off at a sufficient large radius $R_0$, where the Neumann boundary conditions
are imposed as~\footnote{In practice, the value of $R_0$ is sufficiently large compared to the vortex size such that the intrinsic dynamics of composite vortices is not affected.}
\begin{equation}
    \partial_r\psi_i|_{r=R_0}=0,\quad \partial_rA_t|_{r=R_0}=0,\quad\partial_rA_\theta|_{r=R_0}=0.
\end{equation}
At the vortex center $r = 0$, we impose the boundary conditions
\begin{equation}
    \begin{aligned}     &\psi_1|_{r=0}=0,\quad\partial_rA_t|_{r=0}=0,\quad\partial_rA_\theta|_{r=0}=0\,,
    \end{aligned}
\end{equation}
together with 
\begin{equation}
  \begin{aligned} 
    \partial_r\psi_2|_{r=0}=0, \text{ if } S_2=0\,,\\
    \psi_2|_{r=0}=0, \text{ if } S_2\ne 0\,.
 \end{aligned}
\end{equation}
where the former case corresponds to the coreless vortex and the latter one is for the cored vortex.

\begin{figure}[htpb]
        \centering
            \includegraphics[width=0.99\linewidth]{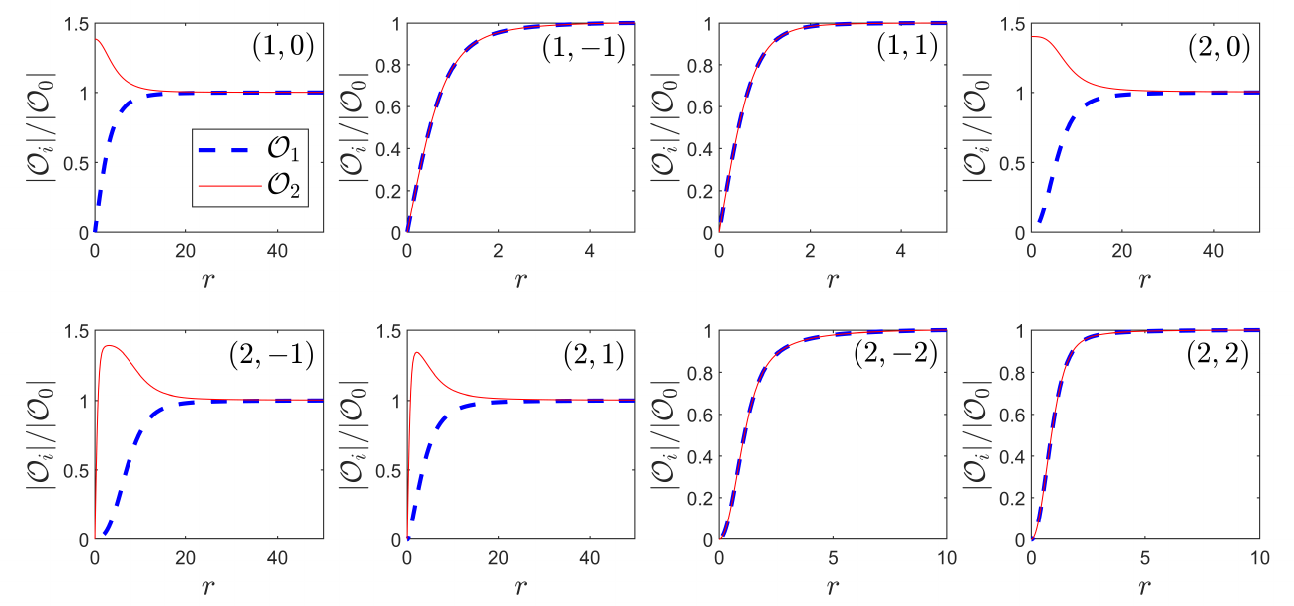}
            \caption{Profile of superfluid condensates of various composite vortices at $T/T_c=0.677$ and $\nu=-0.1$. The first superfluid component is denoted by the blue dashed curves and the second one by the solid red curves. All the condensates are normalized by the condensate value of the first component far from the vortex core $|\mathcal{O}_0|$.}
    \label{profile}
\end{figure}
Then we use the Newton-Raphson method to solve the EoMs and obtain stationary composite vortex configurations for various $S_i$. An representative example of vortex configurations is presented in Figure~\ref{profile}. The profile of a composite vortex depends on the given winding-number pair $(S_1, S_2)$. Some general features can be found. Firstly, the size of vortex for $|S_1|\ne |S_2|$ is much larger than that for $|S_1|=|S_2|$. As we will see later, $|S_1|=|S_2|$ vortices can be dynamical unstable. This implies that the vortex size would grow drastically when they split into the fundamental $(1, 0)$ or $(0,1)$ vortex\,\footnote{The $(0,1)$ vortex can be obtained by choosing $\partial_r\psi_1|_{r=0}=0$ and $\psi_2|_{r=0}=0$.}. Secondly, a bump of the second component would generally grow in the core of composite vortex in the first component when $|S_2|<|S_1|$, see Figures $\ref{profile}(a)$, $\ref{profile}(d)$, $\ref{profile}(e)$ and $\ref{profile}(f)$. This feature could be attributed to the order competing between the two components. At the vortex core of the first component, the second order prevails, while far from the vortex core, two orders compete and suppress each other. Another feature is that, for fixed $|S_1|$ and $|S_2|$, the co-rotating vortex with $S_1S_2>0$ is generally smaller than the counter-rotating vortex with $S_1S_2<0$.

\begin{figure}[htpb]
        \centering
            \includegraphics[width=0.99\linewidth]{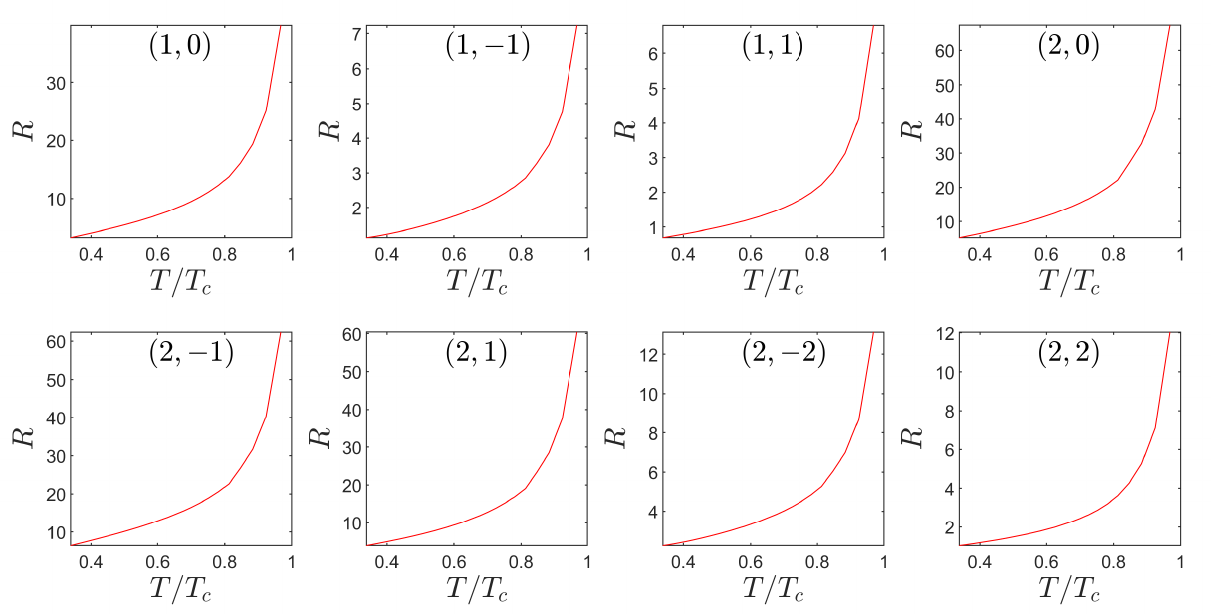}
            \caption{Vortex radius $R$ versus temperature for $\nu=-0.1$. Radius increases as temperature increases for all winding numbers.}
    \label{O_mu}
\end{figure}

\begin{figure}[htpb]
        \centering
            \includegraphics[width=0.98\linewidth]{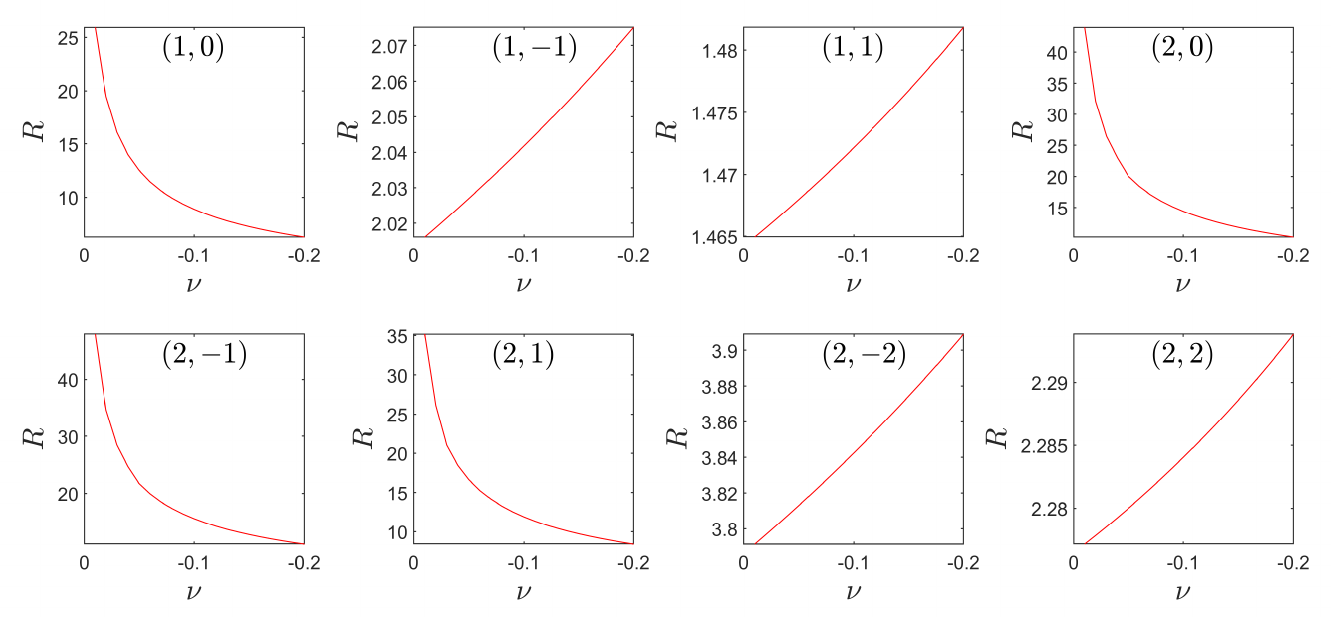}
            \caption{Vortex radius $R$ versus coupling strength $\nu$ at $T/T_c=0.677$. As $\nu$ increases, radius decreases for $|S_1|\ne|S_2|$, while slightly increases for $|S_1|=|S_2|$.}
    \label{O_nu}
\end{figure}

Further insight into the composite vortex can be obtained by considering the vortex radius with respect to the temperature $T$ and the coupling strength $\nu$. In practice, the vortex radius is defined as the size $R$ where $|\mathcal{O}_1(R)|/|\mathcal{O}_0|=0.95$. The temperature dependence of $R$ for various composite vortices is shown in Figure~\ref{O_mu}. It is similar to that of single-component superfluid vortex and vortex-soliton structure in immiscible binary superfluids. As the temperature increases, the vortex radius increases and diverges at the critical temperature $T_c$. The relation between the vortex radius and $\nu$ is a little more complicated, see Figure~\ref{O_nu}. For $|S_1|\ne |S_2|$, the vortex radius decreases as $\nu$ increases, and diverges as $\nu\rightarrow0$. This is in accordance with the fact that the healing length diverges at the critical point, for which we have shown the divergence of healing length at $\nu=0$ from the interface width of the immiscible binary superfluids~\cite{An:2024ebg}. However, for $|S_1|=|S_2|$, the vortex radius increases as $|\nu|$ increases, and the change rate is much smaller than other cases. Moreover, it does not diverge at $\nu=0$. This is due to the fact that there are additional symmetries in~\eqref{eompsi}-\eqref{eomatheta} between the two superfluid components when $|S_1|=|S_2|$. For $S_1=S_2$, the EoMs of the two components are identical. For $S_1=-S_2$, assuming $\psi_1=\psi_2$ would lead to $A_\theta=0$. Again, the EoMs for the two components become identical. For both cases, the full stationary EoMs are equivalent or close to those describing a single component superfluids with a self interaction proportional to $\nu$. Therefore, the vortex size does not diverge even when $\nu=0$, and does not vary much along with $\nu$.

\section{Dynamical instability of composite vortices}
\label{sec4}

A vortex with winding number $|S|>1$ typically has higher energy than a cluster of single-quantum vortices with the same total circulation. Consequently, such multiply quantized vortices will split into single-quantum vortices. For binary superfluids, the additional component makes the situation more complex. In this section, we explore the dynamical instabilities of composite vortices in miscible binary superfluids.

To study the stability properties of a given stationary $(S_1, S_2)$ vortex, we begin with the linear response theory. More precisely, we turn on some perturbations on the stationary background: 
\begin{equation}\label{eq:linear}
    \Phi_i=\Phi_{0i}+\delta\Phi_i,  A_t=A_{t0}+\delta A_t ,  A_r=A_{r0}+\delta A_r ,  A_\theta=A_{\theta 0}+\delta A_\theta\,,
\end{equation}
where $\Phi_{0i}=\psi_{i0} e^{i\Theta+iS_i\theta}$, $A_{t0}$, $A_{r0}$ and $A_{\theta 0}$ are the stationary solutions we get in Section~\ref{sec3}. Thanks to the time translation symmetry and rotation symmetry of the stationary configuration, we can express the perturbations as 
    \begin{equation}\label{eq:pertub}
    \begin{aligned}
        &\delta\Phi_i=u_i(z,r)e^{-i(\omega t-p\theta)}e^{iS_i\theta}, \quad \delta\Phi_i^*=v_i(z,r)e^{-i(\omega t-p\theta)}e^{-iS_i\theta},\\&\delta A_t=a_t(z,r)e^{-i(\omega t-p\theta)},\quad\delta A_r=a_r(z,r)e^{-i(\omega t-p\theta)}, \quad\delta A_\theta=a_\theta(z,r)e^{-i(\omega t-p\theta)}\,,
    \end{aligned}
    \end{equation}
where the integer $p$ specifies the angular momentum of the excitation with respect to the condensate and $\omega$ is the frequency of the excitation.

Plugging~\eqref{eq:linear} into the EoMs and expanding them to the first order of perturbations, we can obtain the linearized equations for perturbations of~\eqref{eq:pertub}. The explicit form of linearized EoMs is the same as that used to study the vortex-bright soliton structure in immiscible binary superfluids~\cite{vortex-soliton}.
This results in schematically a generalized eigenvalue problem:
\begin{equation}
\label{eig}
    M_pu_p=i\omega_p Bu_p, \quad u_p=\{u_1,v_1,u_2,v_2,a_t,a_r,a_\theta\}_p^\mathrm{T}.
\end{equation}
where $M_p$ and $B$ are $7\times 7$ matrices that are determined by the background configuration of a stationary vortex for given $p$. After imposing the in-going boundary condition at the event horizon and the source-free condition at the AdS boundary, we can obtain the spectrum $\omega$ by solving the system numerically. 

Due to the dissipation of the system, the frequency $\omega$ typically takes a complex value, known as quasinormal modes (QNMs) in black hole physics. From the QNMs of the dual black hole, one can obtain both the hydrodynamic modes and non-hydrodynamic of the boundary systems. Since $\partial_t\delta\Phi_i=-i\omega\delta\Phi_i$, the stationary configuration would become dynamical unstable whenever $\mathrm{Im}(\omega)>0$, and $\mathrm{Im}(\omega)$ can characterize the instability strength. As found in~\cite{vortex-soliton}, the complex conjugates of~\eqref{eig} can be obtained by
 \begin{equation}
     p\rightarrow -p,\quad \omega\rightarrow -\omega^*, \quad u_i\leftrightarrow v^*_i ,\quad a_\mu \rightarrow a_\mu^*\,.
 \end{equation}
It follows that whenever $\omega$ is an QNM for given $p$, $-\omega^*$ is an QNM for $-p$. Therefore, without loss of generality, we only need to consider the spectrum with positive $p$. 
\begin{figure}[htpb]
        \centering
            \includegraphics[width=0.9\linewidth]{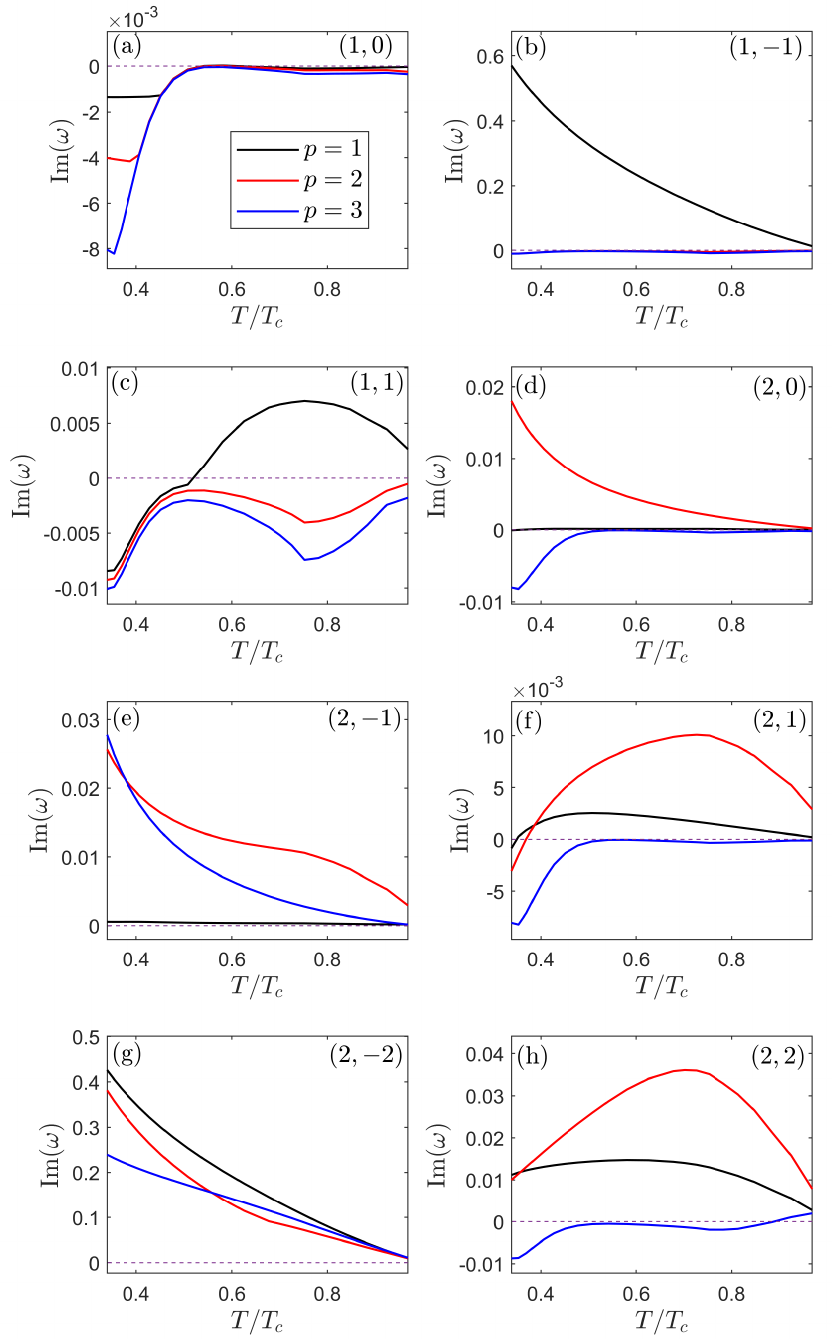}
            \caption{Imaginary part of the excitation frequencies of various composite vortices as a function of temperature for $\nu=-0.1$. The black, red and blue curves correspond to the excitation channels with $p=1$, $p=2$ and $p=3$, respectively.}
    \label{omega_T}
\end{figure}
\begin{figure}[htpb]
        \centering
            \includegraphics[width=0.99\linewidth]{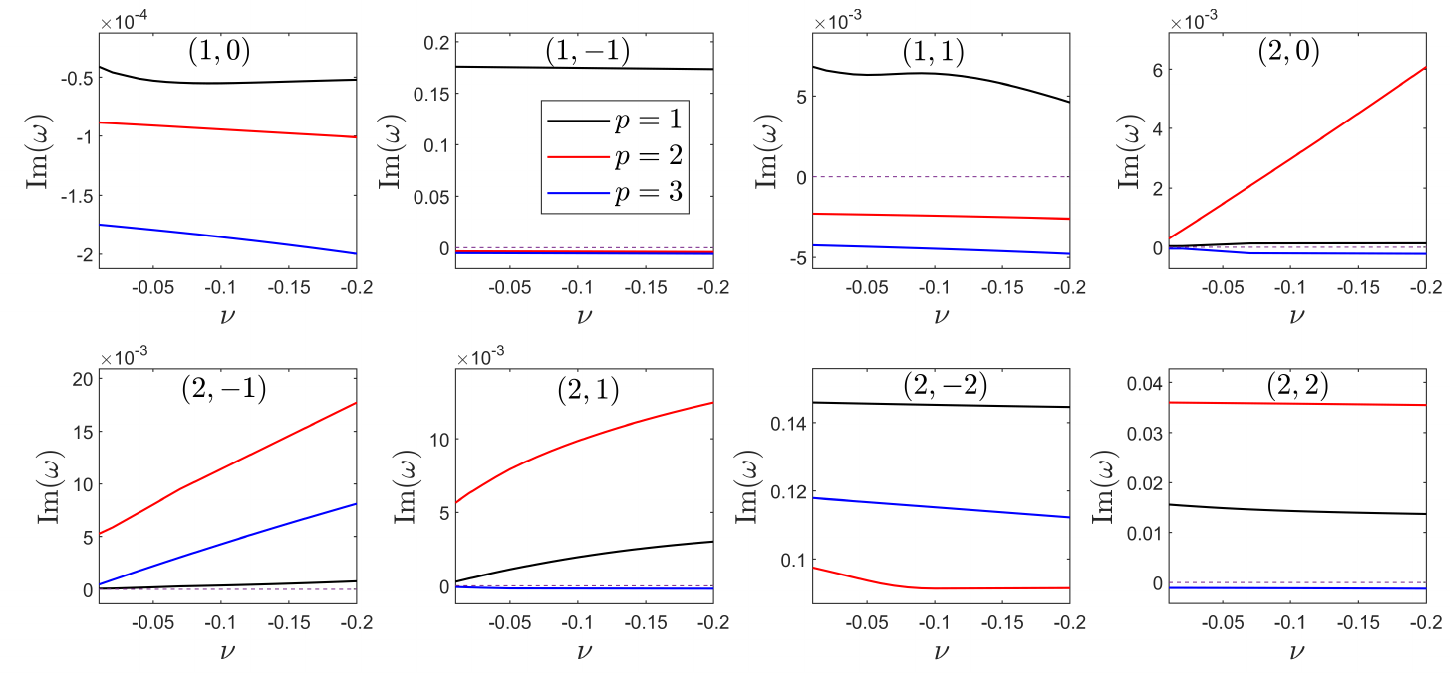}
            \caption{Imaginary part of the excitation frequencies of various composite vortices versus the coupling strength $\nu$ at $T/T_c=0.677$. The curves with different colors correspond to different excitation channels.}
    \label{omega_nu}
\end{figure}

We now present the QNMs spectrum of singly and doubly quantized composite vortices. More precisely, we show how the instability strength $\mathrm{Im}(\omega)$ varies with the temperature in Figure~\ref{omega_T} and the coupling strength $\nu$ in Figure~\ref{omega_nu}. In principle we should consider all integer values of $p$, numerical evidence indicates that only $p=1,2,3$ modes are sufficient since other modes are all stable. One can see that instability strength changes much more drastically with temperature than the case by changing the coupling strength. Depending on the angular-momentum quantum number $p$ and the combination of winding numbers $(S_1, S_2)$, rich dynamical transitions exist.  It suggests the 
instability against splitting of the multiply quantized vortex into singly quantized ones. For dynamically unstable multiquantum vortices with $\mathrm{Im}(\omega)>0$, the exponentially growing modes rapidly drive the system away from the linear regime.

To understand the nonlinear regime of these dynamical instabilities, one has to simulate the full dynamics of the unstable
composite vortices by directly integrating the time-dependent bulk EoMs by using full nonlinear time evolution. The EoMs and time evolution schemes are the same as those used in~\cite{vortex-soliton} for immiscible binary superfluids. We present the stability patterns of singly and doubly quantized
composite vortices. The corresponding dynamics with representative examples from the full time-evolution simulations are presented in Figure~\ref{(2,0)}-\ref{(2,-2)}, which reveals intricate long-time decay behaviors not captured by the linearized approach. Below we scrutinize the dynamical instability of each kind of composite vortex, case by case.
\begin{figure}[htpb]
        \centering
            \includegraphics[width=0.99\linewidth]{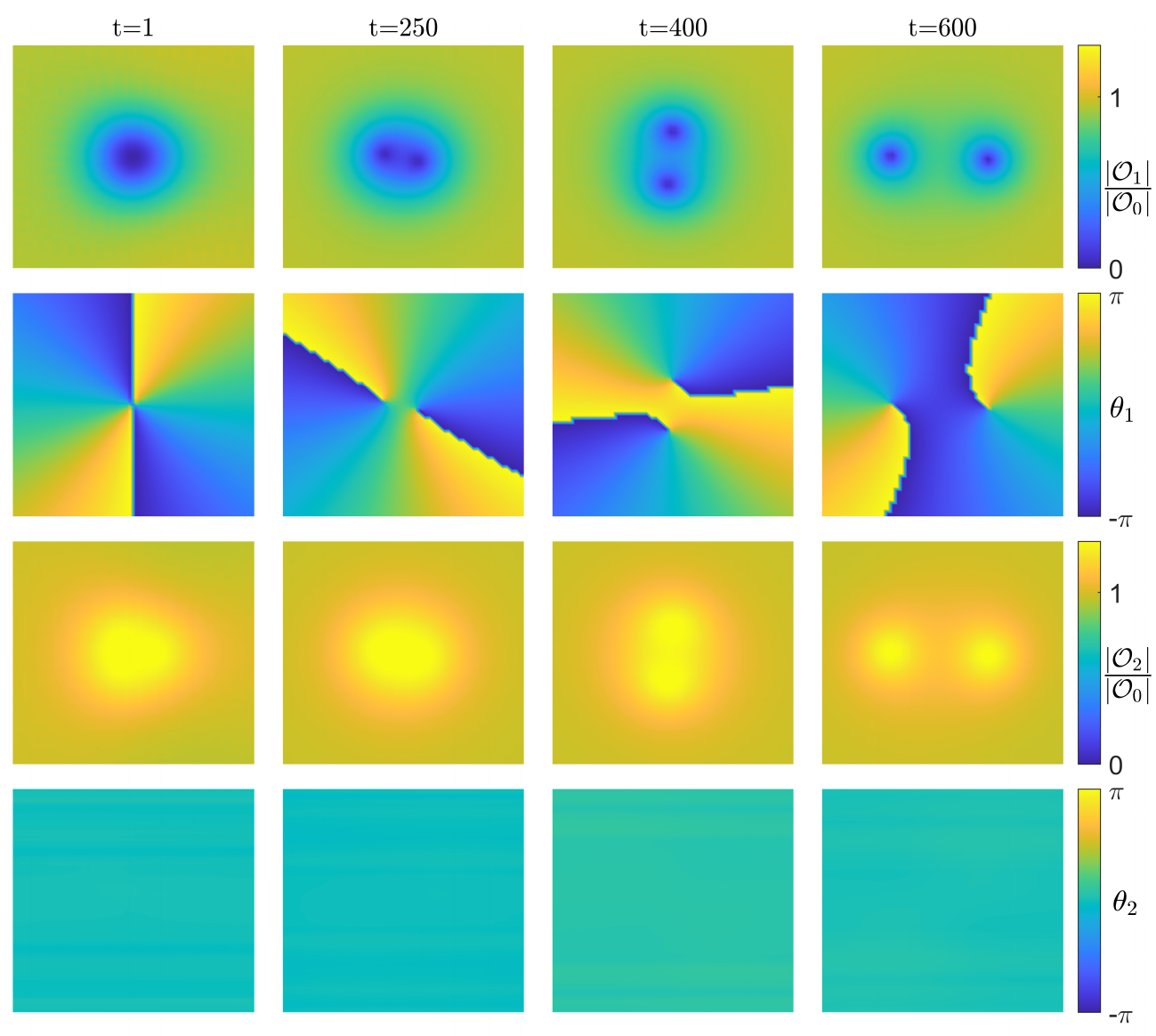}
            \caption{Splitting dynamics for  a slightly perturbed $(2,0)$ vortex at $T/T_c=0.677$ with the inter-component coupling $\nu=-0.2$, for the which dynamical instability only comes from excitations with $|p|=2$. From top to bottom, plotted quantities are $|\mathcal{O}_1|/|\mathcal{O}_0|$, $\theta_1$, $|\mathcal{O}_2|/|\mathcal{O}_0|$ and $\theta_2$, where $\mathcal{O}_0$ is the value of condensate far from all vortices. The size of the plotted region is $25\times 25$. We have fixed $z_h=1$.}
    \label{(2,0)}
\end{figure}
\subsection{$(1,0)\&(2,0)$ vortices}

Firstly, let's consider the case with $S_2=0$, \emph{i.e.} there is no vortex in the second component. Splitting properties of such vortices are similar to those of vortices in single component superfluids.

From the spectrum in Figure~\ref{omega_T} and Figure~\ref{omega_nu}, one can see that all modes of $(1,0)$ vortex are stable, regardless of the temperatures and coupling strength. This is natural since such object is the most elementary constituent of vortices in miscible binary superfluids. There is no channel for such simple object to further split into. On the other hand, winding number is topologically protected, so singly quantized vortex can not disappear unless it is annihilated by another vortex with the opposite winding number $(-1, 0)$. The above argument applies to $(0,\pm 1)$ vortices. Therefore, a singly quantized fundamental vortex is stable. 

For the doubly quantized coreless $(2,0)$ vortex, only the $|p|=2$ mode can be unstable, just as the $S=2$ vortex in single component superfluids~\cite{RN356}. The instability strength of this mode decreases as the temperature increases, tends to zero as the temperature approaches the critical temperature, and slowly increases as the coupling strength $|\nu|$ increases. See Figure~\ref{omega_T}(d). The nonlinear time evolution of this dynamical instability is presented in Figure~\ref{(2,0)}. This is also similar to the single component case. This twofold-symmetric, linear-chain splitting instability results in the splitting into two $(1,0)$ vortices. More precisely, the two-quantum vortex in component 1 splits
into two separated single-quantum vortices. The two $(1,0)$ vortices continue to orbit with each other counter-clockwisely.

\subsection{$(1,\pm 1)$ vortices}

Next, we consider the singly
quantized composite vortices with $S_1=|S_2|=1$. We put them together because they share some similarities. In both cases, only the $p=1$ mode can be unstable. This unstable mode comes from interaction between the two components in binary superfluids and is absent in the axisymmetric solitary vortex with winding number $S=1$ of the single component superfluids. And in both cases, instability strength barely changes as $\nu$ changes. See Figure~\ref{omega_T}(b)$\&$(c) and Figure~\ref{omega_T}. These instabilities describe the splitting of the $(1, \pm1)$ composite vortex into a $(1,0)$ and a $(0,\pm 1)$ vortex. 

In Figure~\ref{(1,-1)}, we show the nonlinear evolution of a slightly perturbed $(1,-1)$ vortex. We see when $p=1$ mode dominates the instability, nonlinear stage of dynamical instability is the decay of the vortiex in to a $(1,0)$ and a $(0, -1)$ vortex. The resulting $(1,0)$ and $(0, -1)$ vortices do not orbit each other since there is only one vortex in each component. And therefore they do not experience from other vortex the Magnus force that gives them tangential velocity. Instead, the two vertices drift away from the original center. This is the typical nonlinear stage of $p=1$ instability\,\footnote{In general, modes with $p>1$ have $Z_p$ symmetry around the center ($r=0$). This feature does not exist for the $p=1$ mode.}. The decay of the $(1,1)$ vortex is similar, but the evolution is much slower since the instability strength is much smaller (see Figure~\ref{omega_T}(c)). 
\begin{figure}[htpb]
        \centering
            \includegraphics[width=0.99\linewidth]{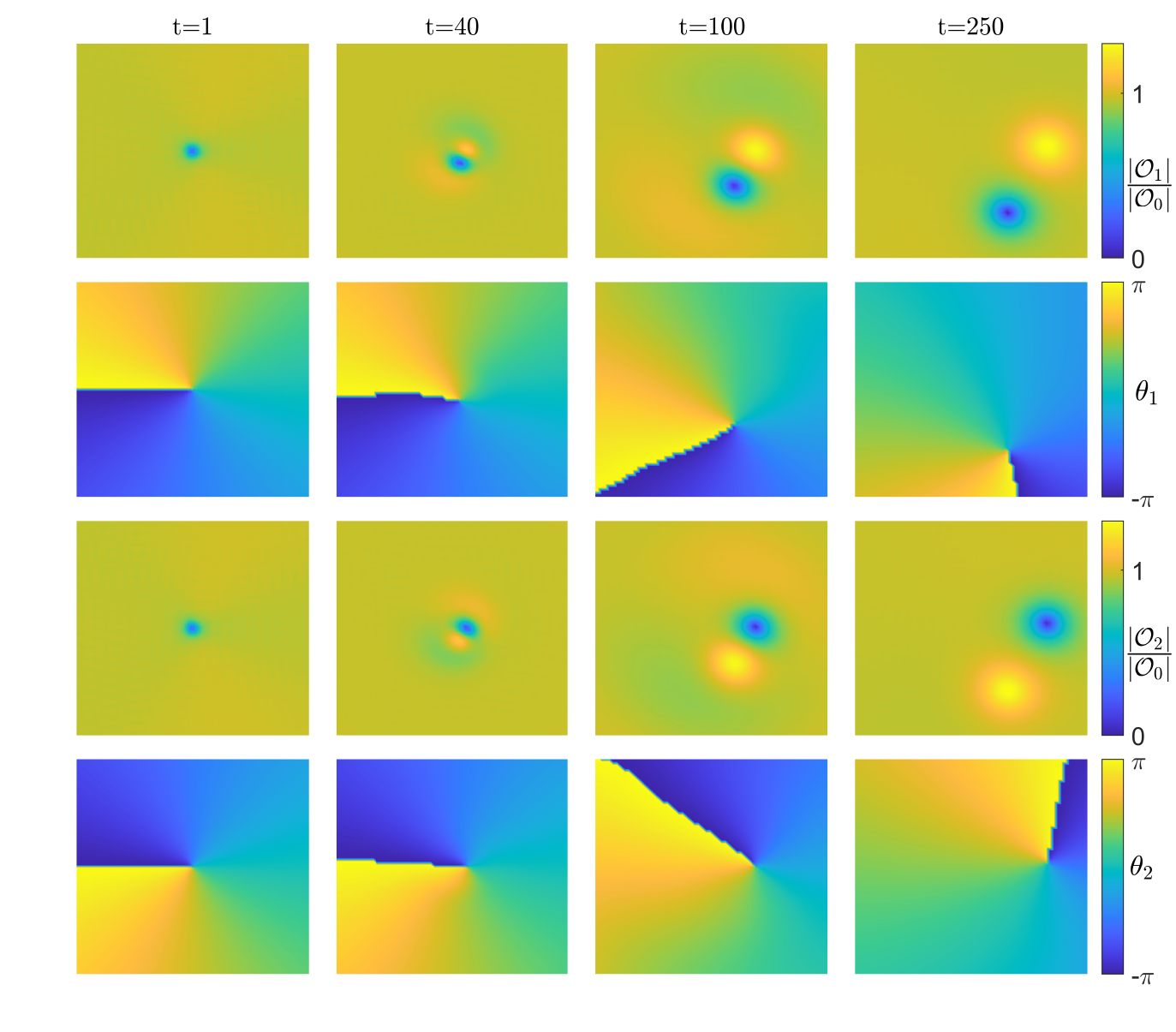}
            \caption{Splitting dynamics for  a slightly perturbed $(1,-1)$ vortex at $T/T_c=0.677$ and $\nu=-0.2$, for which the dynamical instability only comes from excitations with $|p|=1$. Splitting dynamics for $(1,1)$ vortex is similar to this, but evolves much slower. The size of the plotted region is $30\times 30$ with $z_h=1$.}
    \label{(1,-1)}
\end{figure}

Nevertheless, it's worth noting that there is a dynamical transition for $(1,1)$ vortex by changing the temperature. One can see from Figure~\ref{omega_T}(c) that, as the temperature decreases, the instability strength of the $(1,1)$ vortex first increases and then decreases, and below a critical temperature the $(1,1)$ vortex becomes stable. Suppose we generate a pair of a $(1,0)$ and a $(0,1)$ vortex and cool down the temperature. One anticipates that the two vortices would merge into a single $(1,1)$ vortex if they get close enough. In contrast, the $(1,-1)$ vortex is unstable at all temperatures and thus one can not see a stable $(1,-1)$ vortex from merging of a $(1,0)$ and a $(0,-1)$ vortex. In this sense, the effective short range interaction between a $(1,0)$ vortex and a $(0,1)$ vortex can be considered as repulsion at high temperatures and attraction at low temperatures, while it's always repulsion between a $(1,0)$ vortex and a $(0,-1)$ vortex. In fact, we have also verified by numerics that when a $(1,0)$ vortex and a $(0,1)$ vortex come very close at a low temperature, they can merge and form a single $(1,1)$ vortex, just like an inverse process of Figure~\ref{(1,-1)}.

\subsection{$(2,\pm 1)$ vortices}
Now we turn to composite vortices with $S_1=2$ and $|S_2|=1$. Although the absolute values of winding numbers are the same for both cases, their dynamical instability behaviors are quite different.
\begin{figure}[htpb]
        \centering            \includegraphics[width=0.99\linewidth]{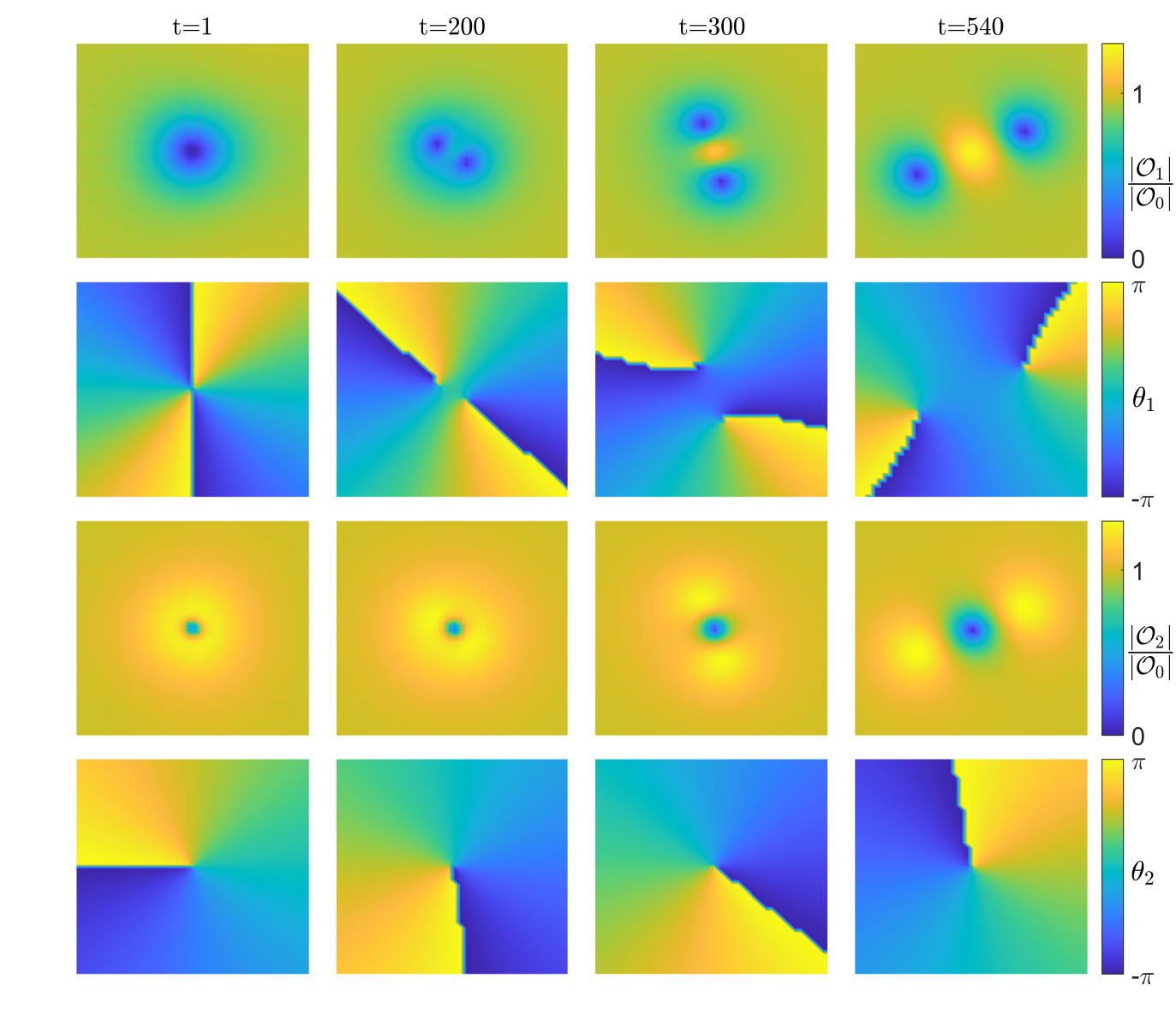}
            \caption{Splitting dynamics for  a slightly perturbed $(2,1)$ vortex at $T/T_c=0.677$ and $\nu=-0.2$, for which the dominant
contribution to the dynamical instability comes from excitations with $|p|=2$. From top to bottom, plotted values are $|\mathcal{O}_1|/|\mathcal{O}_0|$, $\theta_1$, $|\mathcal{O}_2|/|\mathcal{O}_0|$ and $\theta_2$, where $|\mathcal{O}_0|$ is the value of order parameters far from the vortex core. The size of the plotted region is $20\times 20$ with $z_h=1$.  }
    \label{(2,1)}
\end{figure}

For the $(2,1)$ vortex, the unstable modes come from the $p=1,2$ excitations, see Figure~\ref{omega_T}(f). The instability strength of both modes first increases and then decreases as the temperature increases. For sufficiently low temperatures, the dynamical instability disappears. There exist two dynamical transitions for the $(2,1)$ vortex. At high temperatures, the leading unstable mode is the $p=2$ excitation (red curve of Figure~\ref{omega_T}(f)). As the temperature decreases, the dominant unstable mode switches to $p=1$ channel (black curve of Figure~\ref{omega_T}(f)), which is the first transition. And as the temperature goes lower, the second transition happens, for which both modes become stable below a critical temperature. Since $(2,0)$ and $(2,-1)$ vortices are always unstable, this means $(0,1)$ vortex might be acting as a stablizer in the vortex core of $(2,0)$ vortex at low temperature, similar to the vortex-bright soliton structure in immiscible binary superfluids~\cite{vortex-soliton}.

In Figure~\ref{(2,1)}, we show the nonlinear evolution of the $(2,1)$ vortex at a temperature where the $p=2$ channel dominates. At the nonlinear stage of the instability,  the doubly charged vortex in
component 1 splits into two singly charged $(1,0)$ vortices, which orbit a singly
charged vortex in component 2. Therefore, the $(2,1)$ vortex splits directly into two $(1,0)$ vortices and
one $(0,1)$ vortex without intermediate stage.
\begin{figure}[htpb]
        \centering
            \includegraphics[width=0.99\linewidth]{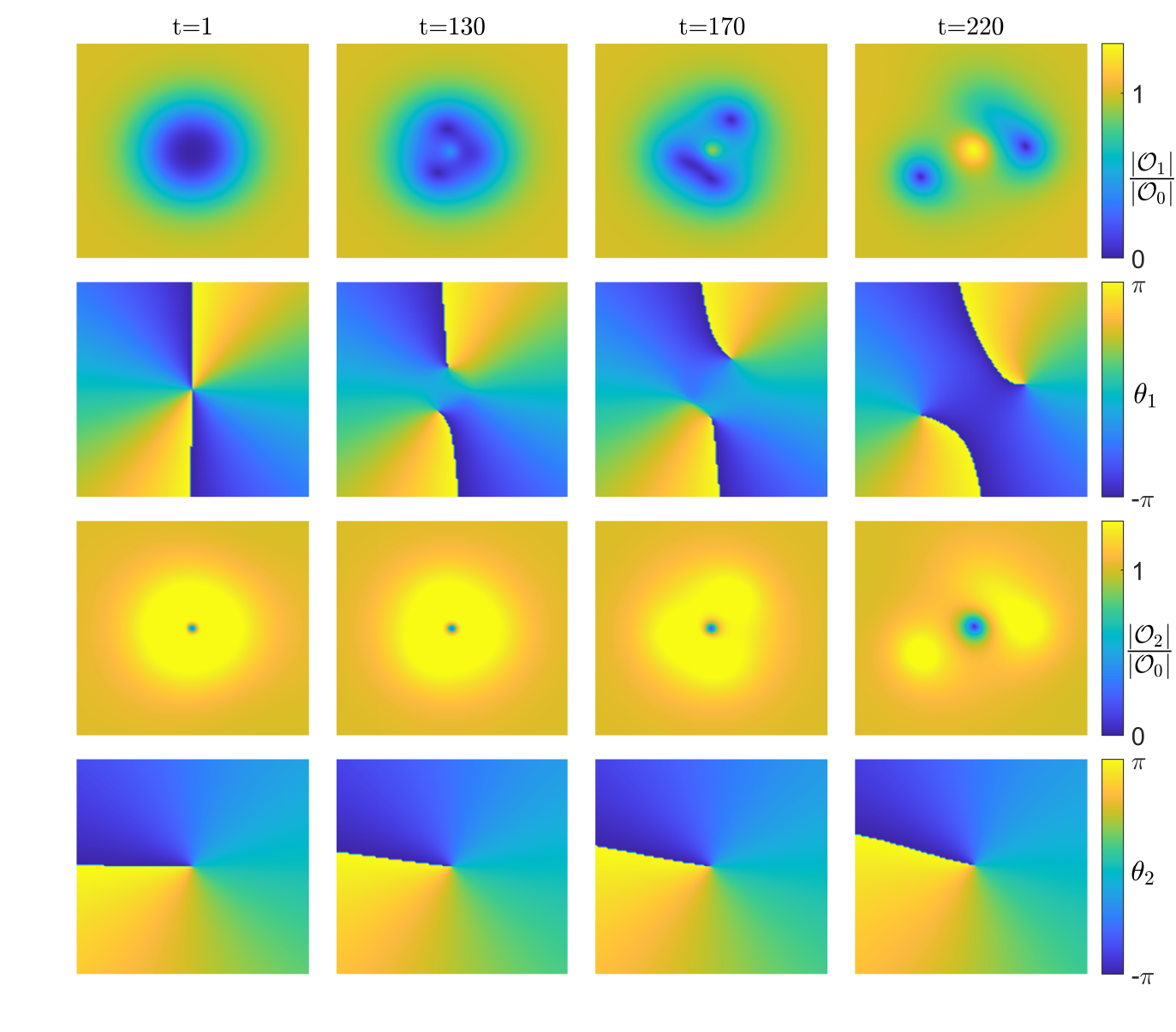}
            \caption{Splitting dynamics for  a slightly perturbed $(2,-1)$ vortex at $T/T_c=0.339$ and $\nu=-0.1$, for which the dominant
contribution to the dynamical instability comes from excitations with $|p|=3$. From top to bottom, plotted values are $|\mathcal{O}_1|/|\mathcal{O}_0|$, $\theta_1$, $|\mathcal{O}_2|/|\mathcal{O}_0|$ and $\theta_2$, where $|\mathcal{O}_0|$ is the value of condensate far from the vortex core. The size of the plotted region is $15\times 15$ with $z_h=1$. }
    \label{(2,-1)}
\end{figure}

For the $(2,-1)$ vortex, as shown in Figure~\ref{omega_T}(e), all $p=1,2,3$ channels are unstable. However, the instability strength of the $p=1$ channel is much smaller compared with other two cases, and thus it never dominates. The instability from the $p=3$ channel might be a consequence of the strongly interacting nature of the system, since the weakly interacting GPE predicts this channel to be stable for $(2,-1)$ vortex~\cite{RN398} (but $p=3$ mode can be unstable for the $(2,-2)$ vortex in GPE). The instability strength of both $p=2$ and $p=3$ excitations decreases as the temperature increases. However, at low temperatures, the $p=3$ channel (red curve) dominates, while at high temperatures, the fastest growing excitation comes from the $p=2$ channel (black curve). When the $p=2$ channel dominates, the nonlinear evolution of $(2,-1)$ vortex is similar to that of the $(2,1)$ vortex in Figure~\ref{(2,1)}. The process is expressed
as the splitting of the $(2,-1)$ vortex into two $(1,0)$ vortices and one $(0,-1)$ vortex.

In Figure~\ref{(2,-1)}, we show a representative example of the nonlinear evolution dominated by the $p=3$ channel. In this case, the excitation with the largest imaginary part is obtained for $|p|=3$. Consequently, one finds an intermediate stage for which the patten has approximately a three-fold symmetry, but no additional vortex is formed. At later times, the doubly charged vortex in component 1 splits into
two singly charged coreless vortices, together with the singly charged antivortex in component 2.

\begin{figure}[htpb]
        \centering
            \includegraphics[width=0.99\linewidth]{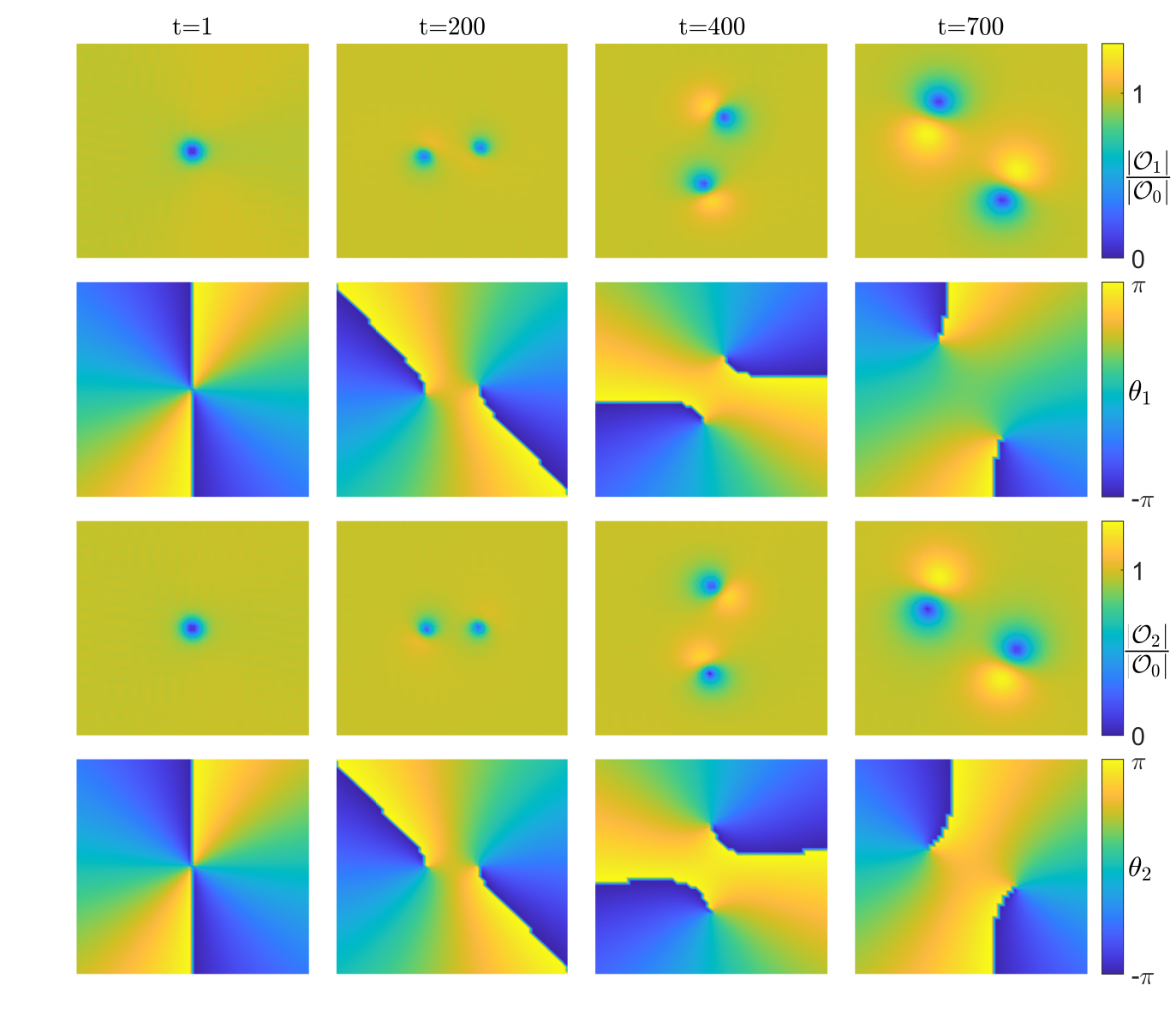}
            \caption{Splitting dynamics for a slightly perturbed $(2,2)$ with $T/T_c=0.677$ and $\nu=-0.2$. The fastest growing mode in the present case is from the channel with $|p|=2$. From top to bottom, plotted values are $|\mathcal{O}_1|/|\mathcal{O}_0|$, $\theta_1$, $|\mathcal{O}_2|/|\mathcal{O}_0|$ and $\theta_2$, where $|\mathcal{O}_0|$ is the value of order parameters far from the vortex core. The size of the plotted region is $25\times 25$ with $z_h=1$. }
    \label{(2,2)}
\end{figure}

\subsection{$(2,\pm 2)$ vortices}

Lastly, we consider $S_1=|S_2|=2$. For both cases, $p=1,2,3$ modes can all be unstable. The temperature dependence of instability strength of these cases are similar to those of $|S_2|=1$. For the $(2,-2)$ vortex, the instability strength decreases monotonically as the temperature increases, while for the $(2,2)$ vortex, the relation is non-monotonic. Let's discuss the unstable modes for each case in more detail.

For the $(2,2)$ vortex, in most parameter space, especially at low temperatures, only $p=1,2$ modes are unstable, see Figure~\ref{omega_T}(h). This is in consistence with zero-temperature weakly interacting results~\cite{RN398}. However, near the critical temperature $T_c$, there is a narrow window of dynamical stability for  the $p=3$ channel. Nevertheless, this excitation does not dominate the system, unless the initial conditions are finely tuned. At high temperatures, the dominant
contribution to the dynamical instability comes from excitations with $|p|=2$. In Figure~\ref{(2,2)}, we show a typical example of the dynamical evolution in this parameter region. The splitting process contains two stages. Firstly, the $(2,2)$ vortex splits into two $(1,1)$ vortices, and they orbit each other counter-clockwisely for a while. Then each $(1,1)$ vortex splits further into two singly quantized vortices. At low temperatures, a dynamical transition occurs and the $p=1$ mode comes to dominate. In this case, the $(2,2)$ vortex would split into a $(2,0)$ vortex and a $(0,2)$ vortex first and then they split into singly quantized vortices separately.
\begin{figure}[htpb]
        \centering
            \includegraphics[width=0.99\linewidth]{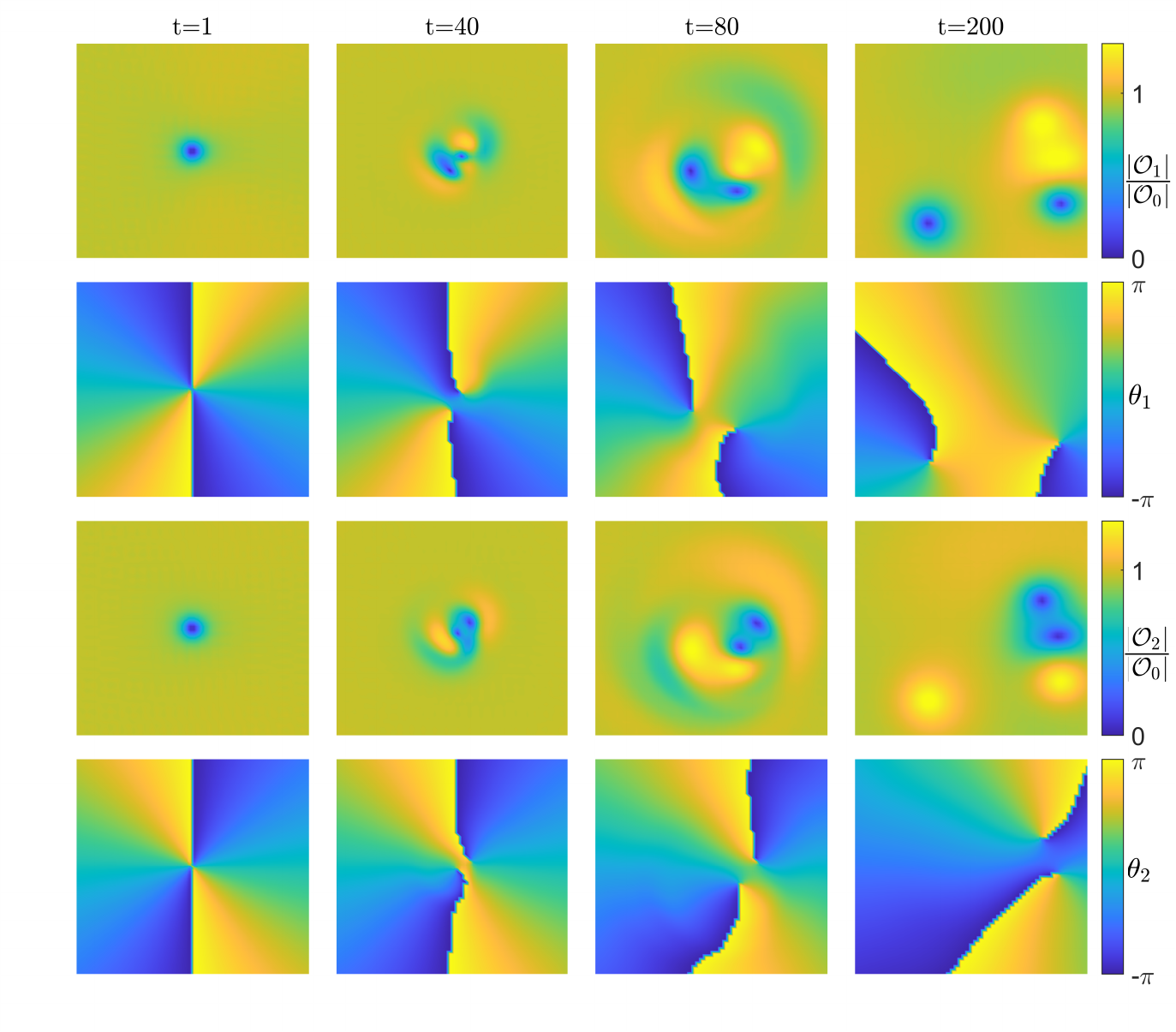}
            \caption{Splitting dynamics for a slightly perturbed $(2,-2)$ vortex at $T/T_c=0.677$ and $\nu=-0.2$. The leading unstable mode is from the channel with $|p|=1$. From top to bottom, plotted values are $|\mathcal{O}_1|/|\mathcal{O}_0|$, $\theta_1$, $|\mathcal{O}_2|/|\mathcal{O}_0|$ and $\theta_2$, where $|\mathcal{O}_0|$ is the value of condensate far from the vortex core. The size of the plotted region is $30\times 30$ with $z_h=1$.}
    \label{(2,-2)}
\end{figure}

For the $(2,-2)$ vortex, as shown in Figure~\ref{omega_T}(g), the excitation with $|p|=1$ dominates the instability for all parameters, although the excitation frequencies for other channels also take positive imaginary parts. This result contrasts with GPE results~\cite{RN398,RN397}, where the $p=2,3$ channel can also dominate, and especially additional vortices are produced when the $p=3$ channel dominates. Figure~\ref{(2,-2)} depicts the decay of the $(2,-2)$ vortex. The splitting pattern is less regular than other cases since all three unstable modes grow together. We do not see clear intermediate stage during this process. The $(2,-2)$ vortex seems to directly split into two $(1,0)$ and two $(0,-1)$ vortices. 

Note that results from GPE show additional vortex formation during the time evolution~\cite{RN394,RN398,RN397}, for example in the $(2,-2)$ vortex subjected to the exotic instability mode $p=3$. This distinction might be due to the strong dissipation effect in holographic superfluids, forcing vortex and anti-vortex to annihilate quickly. In Figure~\ref{(2,-2)_p=3} we manually enhance the $p=3$ perturbation. At intermediate stage, additional vortices forms, but they annihilate shortly. And again, only  two $(1,0)$ and two $(0,-1)$ vortices are left in the end. In all cases we have studied, we don't see the formation of additional long living vortex during the time evolution of a slightly perturbed composite vortex, which could be due to strong dissipation. This result highlights the difference between the vortex dynamics of weakly interacting binary superfluid without dissipation and the strongly interacting case with strong dissipation. 
Moreover, except for the fundamental $(\pm 1, 0)$ and $(0, \pm 1)$ vortices, other composite vortices are dynamical unstable. For a wide range of intercomponent coupling strengths, each splitting processe is the dominant decay mechanism of
the respective stationary composite vortex. As a consequence, the final states are generally singly quantized fundamental vortices.
\begin{figure}[htpb]
        \centering
            \includegraphics[width=0.99\linewidth]{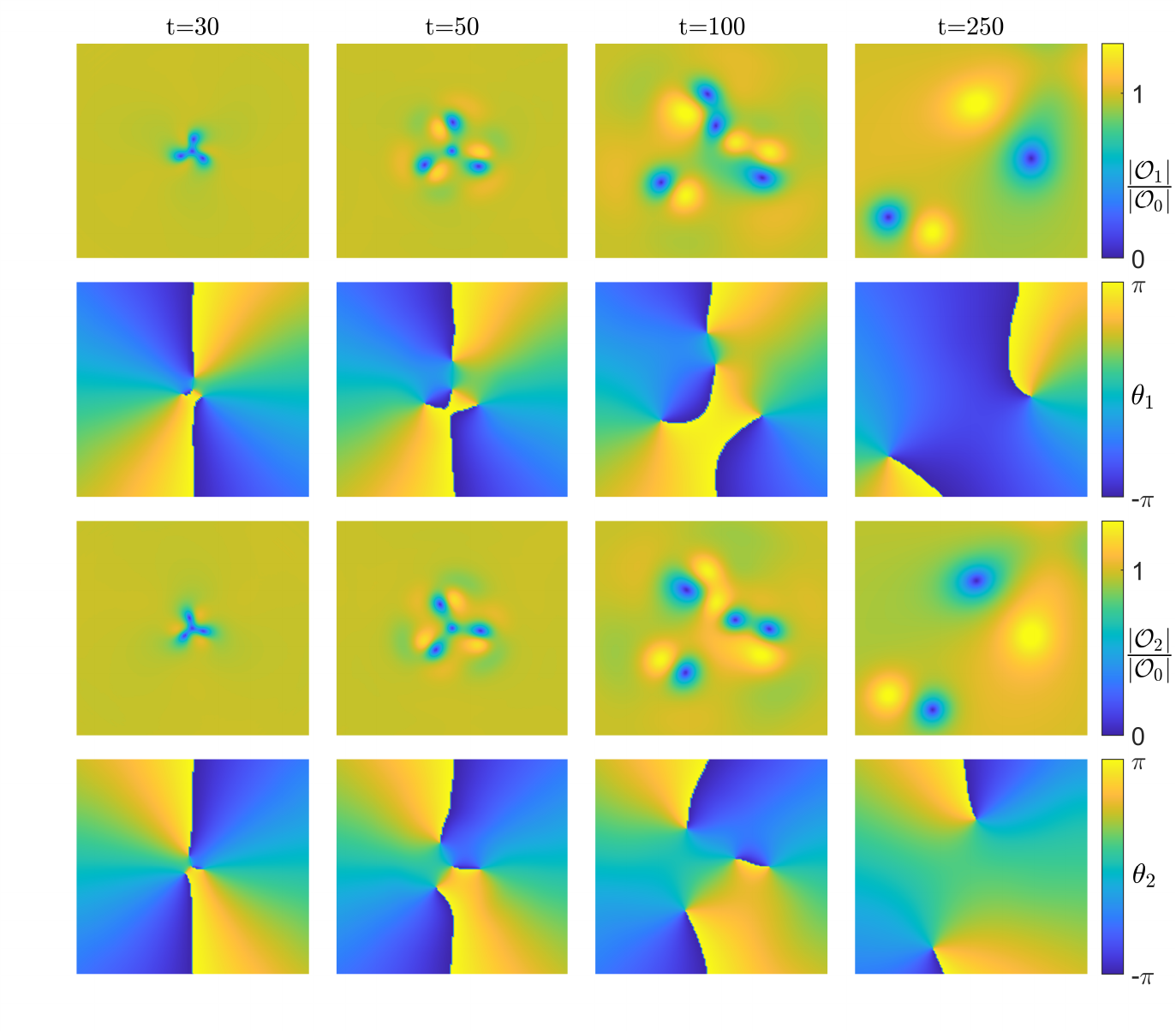}
            \caption{Time evolution of a slightly perturbed $(2,-2)$ vortex with $p=3$ mode manually enhanced at $T/T_c=0.677$ and $\nu=-0.2$. The perturbation added to the background only contains the excitation with $p=3$. There are additional vortices
developing at the intermediate stage.
            From top to bottom, plotted values are $|\mathcal{O}_1|/|\mathcal{O}_0|$, $\theta_1$, $|\mathcal{O}_2|/|\mathcal{O}_0|$ and $\theta_2$, where $|\mathcal{O}_0|$ is the value of order parameters far from the vortex core. The size of the plotted region is $35\times 35$ with $z_h=1$.}
    \label{(2,-2)_p=3}
\end{figure}

\section{Conclusion and discussion}
\label{sec5}

In this work, we have studied the splitting of singly and doubly quantized composite vortices in miscible strongly interacting binary superfluids. To incorporate the temperature and dissipation effects, we take advantage of the holographic duality by mapping the system into gravitational dynamics in AdS spacetime with one higher dimension. Properties of stationary configurations of composite vortices were analyzed, and dynamical instabilities of composite vortices were studied by both linear response theory and nonlinear time evolution. In particular, the splitting behaviors associated with different multipolarities $p$ were then demonstrated by solving the full-time evolution for slightly perturbed composite vortices.

Several decay modes stemming from the multicomponent nature together with finite temperature and dissipation effects were discovered. Turning on one more superfluid component makes the phenomenon much richer than that of the single component superfluids. Vortices in two-component superfluids can share the same vortex core, while being either co-rotating or counter-rotating. Addtional unstable modes appear and rich dynamical behaviors are observed. In our model, finite temperature and dissipation effects are naturally incorporated, making differences to dynamics of the composite vortices compared to the weakly interacting GPE results at zero temperature and dissipation. Several temperature dependent dynamical transitions were identified for various composite vortex configurations, including $(1,1)$, $(2,\pm 1)$ and $(2,2)$ vortices. Except for the fundamental vortices, other composite vortices are dynamical unstable.

We found the final states of all composite vortices are generally singly quantized vortices under generic perturbations, and no additional long living vortex is formed due to strong dissipation in holographic superfluids. To be explicit, the final state of $(S_1,S_2)$ vortex is $S_1$ $(1,0)$ vortices and $|S_2|$ $(0,\pm 1)$ vortices, where the sign depends on the sign of $S_2$. For example, regardless the initial perturbations, the $(2,-2)$ vortex is observed to split into two $(1,0)$ vortices and two $(0,-1)$ vortices, see Figures~\ref{(2,-2)} and~\ref{(2,-2)_p=3}.
This result is in contrast with the GPE results, where additionally formed vortices tend to live much longer due to zero dissipation, and at the late time of evolution the number of vortices could be larger than $|S_1|$ or $|S_2|$~\cite{RN394,RN398,RN397}.\,\footnote{For example, the dynamical instability from GPE shows that the $(2,-2)$ vortex is  unstable against splitting into a $(-1,1)$ vortex, three $(1, 0)$ vortices and three $(0, -1)$ vortices~\cite{RN398}.} Besides, we also found the leading unstable mode can be in disagreement with GPE results, highlighting the difference between the dynamics of composite vortices in strong-interacting and weak-interacting binary superfluids. 
In particular, the leading unstable modes as well as the strength of dynamical instabilities were found to be sensitive to the temperature, as shown in Figure~\ref{omega_T}. Those behaviors hold for a wide range of intercomponent interaction strengths and should be amenable to experimental detection.

It's worth noting that the $(1,1)$ vortex is stable at low temperatures, while the $(1,-1)$ vortex is always unstable. This can be interpreted as that the effective short range interaction can be attractive between a $(1,0)$ vortex and a $(0,1)$ vortex at low temperatures, but always repulsive between a $(1,0)$ vortex and a $(0,-1)$ vortex. This leaves open the possibility of the formation of large scale structures like vortex clusters due to the attraction between co-rotating vortices and the repulsion between counter-rotating vortices in the presence of two superfluid components. While we have limited ourselves to the splitting dynamics and underlying dynamical instabilities of axisymmetric singly and doubly quantized
composite vortices, it would be interesting to consider quantum vortices with higher winding numbers for which one expects more interesting phenomena about the splitting patterns. Besides, there are many more exotic vortex configurations in binary superfluids, such as square vortex
lattices~\cite{PhysRevLett.88.180403,PhysRevLett.93.210403,PhysRevA.85.043613} and vortex sheets~\cite{PhysRevA.79.023606}. It would also be interesting to investigate these structures within holographic binary superfluids incorporating finite temperature and dissipation effects. It is straightforward to generalize the present study to the case with one higher dimension, for which the splitting-induced
intertwining of vortices and other dynamics
of composite defects can develop.

\acknowledgments

This work was partly supported by the National Natural Science Foundation of China Grants No.\,12075298, No.\,12122513, No.\,11991052 and No.\,12047503. We acknowledge the use of the High Performance Cluster at Institute of Theoretical Physics, Chinese Academy of Sciences.




\bibliographystyle{JHEP}
\bibliography{biblio.bib}

\providecommand{\href}[2]{#2}\begingroup\raggedright\begin{thebibliography}{10}

\bibitem{PhysRevLett.101.031601}
S.A.~Hartnoll, C.P.~Herzog and G.T.~Horowitz, \emph{Building a holographic superconductor}, \href{https://doi.org/10.1103/PhysRevLett.101.031601}{\emph{Phys. Rev. Lett.} {\bfseries 101} (2008) 031601}.

\bibitem{Herzog:2008he}
C.P.~Herzog, P.K.~Kovtun and D.T.~Son, \emph{{Holographic model of superfluidity}}, \href{https://doi.org/10.1103/PhysRevD.79.066002}{\emph{Phys. Rev. D} {\bfseries 79} (2009) 066002} [\href{https://arxiv.org/abs/0809.4870}{{\ttfamily 0809.4870}}].

\bibitem{RN400}
Y.~Kawaguchi and T.~Ohmi, \emph{Splitting instability of a multiply charged vortex in a bose-einstein condensate}, \href{https://doi.org/10.1103/PhysRevA.70.043610}{\emph{Physical Review A} {\bfseries 70} (2004) 043610}.

\bibitem{RN399}
P.~Kuopanportti, E.~Lundh, J.A.M.~Huhtamäki, V.~Pietilä and M.~Möttönen, \emph{Core sizes and dynamical instabilities of giant vortices in dilute bose-einstein condensates}, \href{https://doi.org/10.1103/PhysRevA.81.023603}{\emph{Physical Review A} {\bfseries 81} (2010) 023603}.

\bibitem{RN352}
S.~Middelkamp, P.G.~Kevrekidis, D.J.~Frantzeskakis, R.~Carretero-González and P.~Schmelcher, \emph{Bifurcations, stability, and dynamics of multiple matter-wave vortex states}, \href{https://doi.org/10.1103/PhysRevA.82.013646}{\emph{Physical Review A} {\bfseries 82} (2010) 013646}.

\bibitem{RN401}
J.~Räbinä, P.~Kuopanportti, M.I.~Kivioja, M.~Möttönen and T.~Rossi, \emph{Three-dimensional splitting dynamics of giant vortices in bose-einstein condensates}, \href{https://doi.org/10.1103/PhysRevA.98.023624}{\emph{Physical Review A} {\bfseries 98} (2018) 023624}.

\bibitem{RN359}
C.~Ewerz, A.~Samberg and P.~Wittmer, \emph{Dynamics of a vortex dipole in a holographic superfluid}, \href{https://doi.org/10.1007/JHEP11(2021)199}{\emph{Journal of High Energy Physics} {\bfseries 2021} (2021) 199}.

\bibitem{RN357}
V.~Keränen, E.~Keski-Vakkuri, S.~Nowling and K.P.~Yogendran, \emph{Inhomogeneous structures in holographic superfluids. ii. vortices}, \href{https://doi.org/10.1103/PhysRevD.81.126012}{\emph{Physical Review D} {\bfseries 81} (2010) 126012}.

\bibitem{RN356}
S.~Lan, X.~Li, J.~Mo, Y.~Tian, Y.-K.~Yan, P.~Yang et~al., \emph{Splitting of doubly quantized vortices in holographic superfluid of finite temperature}, \href{https://doi.org/10.1007/JHEP05(2023)223}{\emph{Journal of High Energy Physics} {\bfseries 2023} (2023) 223}.

\bibitem{RN393}
S.~Lan, X.~Li, Y.~Tian, P.~Yang and H.~Zhang, \emph{Heating up quadruply quantized vortices: Splitting patterns and dynamical transitions}, \href{https://doi.org/10.1103/PhysRevLett.131.221602}{\emph{Physical Review Letters} {\bfseries 131} (2023) 221602}.

\bibitem{RN391}
J.-H.~Su, C.-Y.~Xia, W.-C.~Yang and H.-B.~Zeng, \emph{Giant vortex in a fast rotating holographic superfluid}, \href{https://doi.org/10.1103/PhysRevD.107.026006}{\emph{Physical Review D} {\bfseries 107} (2023) 026006}.

\bibitem{RN358}
W.-C.~Yang, C.-Y.~Xia, H.-B.~Zeng, M.~Tsubota and J.~Zaanen, \emph{Motion of a superfluid vortex according to holographic quantum dissipation}, \href{https://doi.org/10.1103/PhysRevB.107.144511}{\emph{Physical Review B} {\bfseries 107} (2023) 144511}.

\bibitem{RN349}
G.~Catelani and E.A.~Yuzbashyan, \emph{Coreless vorticity in multicomponent bose and fermi superfluids}, \href{https://doi.org/10.1103/PhysRevA.81.033629}{\emph{Physical Review A} {\bfseries 81} (2010) 033629}.

\bibitem{RN395}
A.~Chaika, A.~Richaud and A.~Yakimenko, \emph{Making ghost vortices visible in two-component bose-einstein condensates}, \href{https://doi.org/10.1103/PhysRevResearch.5.023109}{\emph{Physical Review Research} {\bfseries 5} (2023) 023109}.

\bibitem{RN361}
E.G.~Charalampidis, P.G.~Kevrekidis, D.J.~Frantzeskakis and B.A.~Malomed, \emph{Vortex-soliton complexes in coupled nonlinear schr\"odinger equations with unequal dispersion coefficients}, \href{https://doi.org/10.1103/PhysRevE.94.022207}{\emph{Physical Review E} {\bfseries 94} (2016) 022207}.

\bibitem{RN397}
S.~Ishino, M.~Tsubota and H.~Takeuchi, \emph{Counter-rotating vortices in miscible two-component bose-einstein condensates}, \href{https://doi.org/10.1103/PhysRevA.88.063617}{\emph{Physical Review A} {\bfseries 88} (2013) 063617}.

\bibitem{RN350}
K.J.H.~Law, P.G.~Kevrekidis and L.S.~Tuckerman, \emph{Stable vortex--bright-soliton structures in two-component bose-einstein condensates}, \href{https://doi.org/10.1103/PhysRevLett.105.160405}{\emph{Physical Review Letters} {\bfseries 105} (2010) 160405}.

\bibitem{RN351}
M.~Pola, J.~Stockhofe, P.~Schmelcher and P.G.~Kevrekidis, \emph{Vortex--bright-soliton dipoles: Bifurcations, symmetry breaking, and soliton tunneling in a vortex-induced double well}, \href{https://doi.org/10.1103/PhysRevA.86.053601}{\emph{Physical Review A} {\bfseries 86} (2012) 053601}.

\bibitem{RN365}
V.P.~Ruban, \emph{Instabilities of a filled vortex in a two-component bose–einstein condensate}, \href{https://doi.org/10.1134/S0021364021080117}{\emph{JETP Letters} {\bfseries 113} (2021) 532}.

\bibitem{RN394}
D.V.~Skryabin, \emph{Instabilities of vortices in a binary mixture of trapped bose-einstein condensates: Role of collective excitations with positive and negative energies}, \href{https://doi.org/10.1103/PhysRevA.63.013602}{\emph{Physical Review A} {\bfseries 63} (2000) 013602}.

\bibitem{RN348}
J.~Xing, W.~Bai, B.~Xiong, J.-H.~Zheng and T.~Yang, \emph{Structure and dynamics of binary bose–einstein condensates with vortex phase imprinting}, \href{https://doi.org/10.1007/s11467-023-1316-0}{\emph{Frontiers of Physics} {\bfseries 18} (2023) 62302}.

\bibitem{RN412}
A.~Richaud, V.~Penna and A.L.~Fetter, \emph{Dynamics of massive point vortices in a binary mixture of bose-einstein condensates}, \href{https://doi.org/10.1103/PhysRevA.103.023311}{\emph{Physical Review A} {\bfseries 103} (2021) 023311}.

\bibitem{RN411}
A.~Richaud, V.~Penna, R.~Mayol and M.~Guilleumas, \emph{Vortices with massive cores in a binary mixture of bose-einstein condensates}, \href{https://doi.org/10.1103/PhysRevA.101.013630}{\emph{Physical Review A} {\bfseries 101} (2020) 013630}.

\bibitem{RN398}
P.~Kuopanportti, S.~Bandyopadhyay, A.~Roy and D.~Angom, \emph{Splitting of singly and doubly quantized composite vortices in two-component bose-einstein condensates}, \href{https://doi.org/10.1103/PhysRevA.100.033615}{\emph{Physical Review A} {\bfseries 100} (2019) 033615}.

\bibitem{vortex-soliton}
Y.~An and L.~Li, \emph{{(In)stability of symbiotic vortex-bright soliton in holographic immiscible binary superfluids}},  \href{https://arxiv.org/abs/2409.08310}{{\ttfamily 2409.08310}}.

\bibitem{Basu:2010fa}
P.~Basu, J.~He, A.~Mukherjee, M.~Rozali and H.-H.~Shieh, \emph{{Competing Holographic Orders}}, \href{https://doi.org/10.1007/JHEP10(2010)092}{\emph{JHEP} {\bfseries 10} (2010) 092} [\href{https://arxiv.org/abs/1007.3480}{{\ttfamily 1007.3480}}].

\bibitem{Cai:2013wma}
R.-G.~Cai, L.~Li, L.-F.~Li and Y.-Q.~Wang, \emph{{Competition and Coexistence of Order Parameters in Holographic Multi-Band Superconductors}}, \href{https://doi.org/10.1007/JHEP09(2013)074}{\emph{JHEP} {\bfseries 09} (2013) 074} [\href{https://arxiv.org/abs/1307.2768}{{\ttfamily 1307.2768}}].

\bibitem{Yang:2019ibe}
W.-C.~Yang, C.-Y.~Xia, H.-B.~Zeng and H.-Q.~Zhang, \emph{{Phase Separation and Exotic Vortex Phases in a Two-Species Holographic Superfluid}}, \href{https://doi.org/10.1140/epjc/s10052-021-08838-x}{\emph{Eur. Phys. J. C} {\bfseries 81} (2021) 21} [\href{https://arxiv.org/abs/1907.01918}{{\ttfamily 1907.01918}}].

\bibitem{Yao:2022fov}
S.~Yao, Y.~Tian, P.~Yang and H.~Zhang, \emph{{Baby skyrmion in two-component holographic superfluids}}, \href{https://doi.org/10.1007/JHEP08(2023)055}{\emph{JHEP} {\bfseries 08} (2023) 055} [\href{https://arxiv.org/abs/2210.12490}{{\ttfamily 2210.12490}}].

\bibitem{An:2024ebg}
Y.-P.~An, L.~Li, C.-Y.~Xia and H.-B.~Zeng, \emph{{Interface dynamics of strongly interacting binary superfluids}}, \href{https://doi.org/10.1103/PhysRevD.109.106022}{\emph{Phys. Rev. D} {\bfseries 109} (2024) 106022} [\href{https://arxiv.org/abs/2401.09189}{{\ttfamily 2401.09189}}].

\bibitem{An:2024dkn}
Y.-P.~An, L.~Li and H.-B.~Zeng, \emph{{Quantum analog to flapping of flags: interface instability for co-flow binary superfluids}}, \href{https://doi.org/10.1007/JHEP10(2024)014}{\emph{JHEP} {\bfseries 10} (2024) 014} [\href{https://arxiv.org/abs/2406.13965}{{\ttfamily 2406.13965}}].

\bibitem{An:2024ctq}
Y.~An, B.~Gouteraux and L.~Li, \emph{{Uncovering thermodynamic origin of counterflow and coflow instabilities in miscible binary superfluids}},  \href{https://arxiv.org/abs/2411.01972}{{\ttfamily 2411.01972}}.

\bibitem{PhysRevLett.88.180403}
E.J.~Mueller and T.-L.~Ho, \emph{Two-component bose-einstein condensates with a large number of vortices}, \href{https://doi.org/10.1103/PhysRevLett.88.180403}{\emph{Phys. Rev. Lett.} {\bfseries 88} (2002) 180403}.

\bibitem{PhysRevLett.93.210403}
V.~Schweikhard, I.~Coddington, P.~Engels, S.~Tung and E.A.~Cornell, \emph{Vortex-lattice dynamics in rotating spinor bose-einstein condensates}, \href{https://doi.org/10.1103/PhysRevLett.93.210403}{\emph{Phys. Rev. Lett.} {\bfseries 93} (2004) 210403}.

\bibitem{PhysRevA.85.043613}
P.~Kuopanportti, J.A.M.~Huhtam\"aki and M.~M\"ott\"onen, \emph{Exotic vortex lattices in two-species bose-einstein condensates}, \href{https://doi.org/10.1103/PhysRevA.85.043613}{\emph{Phys. Rev. A} {\bfseries 85} (2012) 043613}.

\bibitem{PhysRevA.79.023606}
K.~Kasamatsu and M.~Tsubota, \emph{Vortex sheet in rotating two-component bose-einstein condensates}, \href{https://doi.org/10.1103/PhysRevA.79.023606}{\emph{Phys. Rev. A} {\bfseries 79} (2009) 023606}.

\end{thebibliography}\endgroup

\end{document}